\documentclass[aoas,MSNbibl,nameyear,seceqn,rotating,dvips]{arximspdf}
\usepackage{dcolumn}
\usepackage{multirow}
\usepackage{graphicx}

\doi{10.1214/13-AOAS663} 
\volume{7}
\issue{4}
\pubyear{2013}
\firstpage{2180}
\lastpage{2204}

\makeatletter

\newcolumntype{d}[1]{D{.}{.}{#1}}

\newproclaim{scenario}{Scenario}

\newtheorem{theorem}{Theorem}

\newcommand{\boldX}{\mathbf{X}}
\newcommand{\boldy}{\mathbf{y}}
\newcommand{\boldu}{\mathbf{u}}
\newcommand{\boldbeta}{\bolds{\beta}}
\newcommand{\boldtheta}{\bolds{\theta}}

\makeatother

\begin{document}
\begin{frontmatter}

\title{A new class of flexible link functions with application to species
co-occurrence in cape floristic region}
\runtitle{New class of flexible link function}

\begin{aug}
\author[A]{\fnms{Xun} \snm{Jiang}\thanksref{m1}\ead[label=e1]{xun.jiang@uconn.edu}},
\author[A]{\fnms{Dipak K.} \snm{Dey}\corref{}\thanksref{m1}\ead[label=e2]{dipak.dey@uconn.edu}},
\author[B]{\fnms{Rachel} \snm{Prunier}\thanksref{m2}\ead[label=e3]{prunierr@wcsu.edu}},\\
\author[C]{\fnms{Adam M.} \snm{Wilson}\thanksref{m3}\ead[label=e4]{adam.wilson@yale.edu}}
\and
\author[D]{\fnms{Kent E.} \snm{Holsinger}\thanksref{m1}\ead[label=e5]{kent.holsinger@uconn.edu}}
\runauthor{X. Jiang et al.}
\affiliation{University of Connecticut\thanksmark{m1},
Western Connecticut State University\thanksmark{m2} and
Yale University\thanksmark{m3}}
\address[A]{X. Jiang\\
D. K. Dey\\
Department of Statistics\\
University of Connecticut \\
215 Glenbrook Rd. Unit 4120 \\
Storrs, Connecticut 06269\\
USA\\
\printead{e1}\\
\hphantom{E-mail: }\printead*{e2}}
\address[B]{R. Prunier\\
Department of Biological\\
\quad and Environmental Sciences\\
Western Connecticut State University\\
181 White Street \\
Danbury, Connecticut 06810\\
USA\\
\printead{e3}}
\address[C]{A. M. Wilson\\
Department of Ecology\\
\quad and Evolutionary Biology \\
Yale University \\
165 Prospect Street \\
New Haven, Connecticut 06520\\
USA \\
\printead{e4}}
\address[D]{K. E. Holsinger\\
Department of Ecology\\
\quad and Evolutionary Biology\\
University of Connecticut \\
75 N. Eagleville Road, Unit 3043\hspace*{26.2pt} \\
Storrs, Connecticut 06269\\
USA\\
\printead{e5}}
\end{aug}

\received{\smonth{12} \syear{2012}}
\revised{\smonth{6} \syear{2013}}

%
\begin{abstract}
Understanding the mechanisms that allow biological species to co-occur
is of great interest to ecologists. Here we investigate the factors
that influence co-occurrence of members of the genus \textit{Protea} in
the Cape Floristic Region of southwestern Africa, a global hot spot of
biodiversity. Due to the binomial nature of our response, a critical
issue is to choose appropriate link functions for the regression model.
In this paper we propose a new family of flexible link functions for
modeling binomial response data. By introducing a power parameter into
the cumulative distribution function (c.d.f.) corresponding to a symmetric
link function and its mirror reflection, greater flexibility in
skewness can be achieved in both positive and negative directions.
Through simulated data sets and analysis of the \textit{Protea}
co-occurrence data, we show that the proposed link function is quite
flexible and performs better against link misspecification than
standard link functions.
\end{abstract}

%

\end{frontmatter}

\section{Introduction}\label{sec1}
Understanding the underlying processes that govern the assembly of
biological communities has long been of great interest to
ecologists. Obviously, in the absence of species interactions and
species habitat preferences, the probability that two species co-occur
in a site would simply be the product of the site occupancy
probabilities for each of the species. In most biological communities,
however, competition [\citet{elton1946competition}] and individual
response to the environment [\citet{weiher1995assembly}] are
likely to
play important roles in determining the species composition of local
communities. Since phenotypic traits of species and environmental
factors could mediate both competition and individual response, the
probability of co-occurrence could also be influenced by both the
traits of the species and the specific environmental
conditions associated with a site. In this study we
investigate the processes of community assembly in a well-defined
clade, the genus \textit{Protea} in the Cape Floristic Region (CFR) of
southwestern Africa. The response variable, the number of
co-occurrences of a certain pair of \textit{Protea} species, arises
naturally as a binomial variable when we define co-occurrence as the
number of sites in which two species co-occur within naturally nested
watersheds. We take into consideration the spatial association among
the co-occurrence of \textit{Protea} species since it is natural to
suspect areas close by would tend to have similar number of
co-occurrences as a result of a latent spatial effect. Our primary
interest in this study is to build a comprehensive model that could
identify important factors influencing the assembly of \textit{Protea}
communities, while adjusting for both spatial association and
prevalence of \textit{Protea} in CFR.

The usual way to model the binomial response is to use a Generalized
Linear Model (GLM), where we model the latent probability of
``success'' by a linear function of covariates through a link function
[\citet{mccullagh1989generalized}]. The logit, probit and Student
$t$ link
functions are three of the common links used in GLM. However, the link
functions mentioned above are ``symmetric'' links in the sense that
they assume that the latent probability of a given binomial response
approaches 0 with the same rate as it approaches 1. Equivalently, the
probability density function (p.d.f.) that corresponds to the inverse
cumulative distribution function (c.d.f.) of the link function is
symmetric. However, this may not be a reasonable assumption in many
cases. A commonly adopted asymmetric link function is the
complementary loglog (cloglog) link function. However, the cloglog
link has a fixed negative skewness. As a result, it lacks both the
flexibility to let the data tell how much skewness should be
incorporated and the ability to allow for positive skewness. In short,
binomial data might often be better modeled with flexible link
functions that allow for both positive and negative skew and that
allow the data to determine the amount of skewness required.

Much work has been done to introduce flexibility into the link
functions. \citet{Aranon1981} proposed two separate one-parameter
models for additional flexibility in the logistic model.
\citet{GuerJohnuse1982} used Box--Cox transformation on the odds
ratio to form a more flexible class of model. \citet{Jonerepl2004}
proposed a family of flexible distributions based on the distribution of
order statistics. \citet{StukStukgene1988} proposed a two-parameter
class of generalized logistic models. Stukel's model approximates many
standard symmetric and asymmetric link functions quite well, but in a
Bayesian framework, it may result in improper posteriors when the usual
improper uniform prior is used in regressions [\citet{ChenDeyShaonew1999}].
\citet{kim2008flexible} proposed a class of generalized skewed $t$-link
models using a latent variable approach, which achieves proper
posteriors for regression coefficients under uniform
priors. Unfortunately, the range of the skewness for generalized
skewed $t$-link is limited due to a constraint on the shape parameter
required for identifiability of the model. More recently,
\citet{WangDeygene2010} propose the generalized extreme value link
function to allow more flexible skewness controlled by the shape
parameter, but the standard logistic and probit links are not among
the special cases of this family.

Several authors have proposed an additional power parameter on the
c.d.f.
corresponding to standard link
functions. \citet{nagler1994scobit} introduces the Scobit model, which
is a generalization of the logistic model by introduction of a
power parameter. In the psychology literature, \citet{Samelogi2000}
proposes the Logistic Positive Exponent Family using similar
ideas. These models are part of the asymmetric parametric family
proposed by \citet{Aranon1981} under some
re-parameterizations. \citet{gupta2008analyzing} propose the power
normal distribution to accommodate skewness and discuss its
advantages over the skew normal distribution. However, even though
those link functions with power parameters can accommodate flexible
skewness in one direction (e.g., positive skewness in the Scobit link),
the skewness in the opposite direction can be asymmetrically limited.

In this paper we propose a new class of symmetric power link functions
to model binary and binomial data, and apply it to the \textit{Protea}
species co-occurrence data. The rest of the paper is organized as
follows. We introduce the \textit{Protea} species co-occurrence data in
Section~\ref{sec2}. In Section~\ref{sec3} we propose a general class
of power link functions based on the c.d.f. corresponding to a symmetric
baseline link function and its mirror reflection. Section~\ref{sec4}
discusses the prior specification and posterior proprieties of the
parameter in the proposed model under a fully Bayesian framework. In
Section~\ref{sec10} we introduce spatial random effects in the model
to account for the spatial association in the co-occurrence
data. Section~\ref{sec5} clarifies some computational issues in the
model as well as the criteria for model comparisons. Several comprehensive
simulation studies are reported in Section~\ref{sec6} with detailed
discussions. Finally, in Section~\ref{sec7} we fit the proposed model
on the \textit{Protea} species co-occurrence data. We conclude our
paper in Section~\ref{sec8} and all the proofs of the theorems are
deferred to the \hyperref[app]{Appendices}.

\section{The \textit{Protea} species co-occurrence data}\label{sec2}

\begin{figure}

\includegraphics{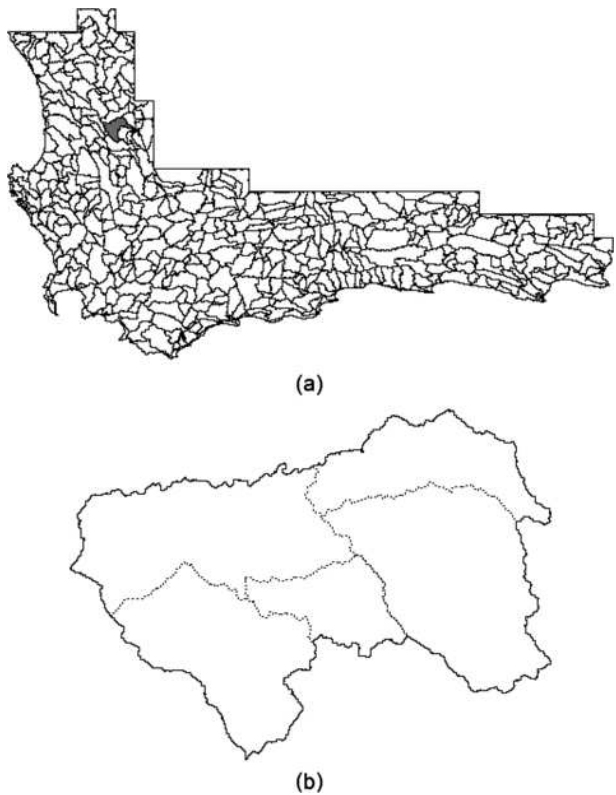}

\caption{Illustration of the watersheds scale in CFR. \textup{(a)} The CFR
showing the boundaries between the large watersheds within which
smaller watersheds are nested. \textup{(b)} Smaller watersheds nested within
one particular parent watershed. This parent watershed is highlighted
in grey in part \textup{(a)}.} \label{fig6}
\end{figure}

The Cape Floristic Region (CFR) is a region with remarkable biological
diversity. The \textit{Protea} species co-occurrence data we study here
is derived from the Protea Atlas data set
(\url{http://www.proteaatlas.org.za}), which includes 96,253 occurrence
records for the 71 species in the genus \textit{Protea} from 44,415
sites. Wilson and Prunier (unpublished data) constructed a series of
nested watersheds covering the CFR using the 3 arc-second (90~m)
research-grade digital elevation model collected by the Shuttle Radar
Topography Mission (available at \url{http://seamless.usgs.gov}) using the
r.watersheds function in GRASS [\citet{GRASS-2008}]. The co-occurrence
data used here correspond to species co-occurrences within watersheds
having a mean area of approximately 55 km$^2$ ($\pm$40 km$^2$)
nested within larger watersheds with a mean area of approximately 540
km$^2$ ($\pm$425 km$^2$) [see Figure~\ref{fig6}(a) for an
illustration of nested watersheds in CFR]. The smaller watersheds are
considerably larger than those usually considered in community
assembly studies [\citet{Vamosi-etal-2009}]. As a result, factors that
are associated with a reduced probability of co-occurrence in this
analysis may reflect either the consequences of competitive
interactions or of habitat segregation among different watersheds.

%
%

%
\begin{figure}

\includegraphics{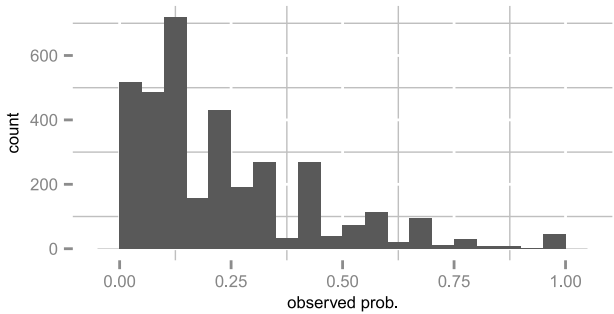}

\caption{Histogram of empirical frequencies of Protea species co-occurence.}
\label{fig7}
\end{figure}

The data are binomial because for each pair of \textit{Protea} species,
we record the number of smaller watersheds at which a particular pair
co-occurs out of the total number of smaller watersheds contained
within each larger watershed. The data are then aggregated across the
larger watersheds to cover the entire CFR. In this study we record
10,256 observations involving ${71 \choose2}$ pairs of \textit{Protea}
species. In Figure~\ref{fig7} we plot the
histogram of observed probability of co-occurrence (number of \textit
{Protea} observed co-occurrences between species pairs divided by
the total number of co-occurrences possible). Although this histogram
is not excessively skewed, it is reasonable to suspect that an
asymmetric link function will be more appropriate for these data,
because the histograms for individual species pairs can be very
skewed. Our observational unit is species pairs, and for each pair of
species we record seven traits that could affect the probability that
they co-occur. They are as follows: (1) phylogenetic distance, which is
proportional to the time since the two species diverged from a common
ancestor, (2) whether or not they differ in fire survival strategy,
(3) the difference in plant height, (4) the difference in month of
maximum flowering, (5) whether or not they share a pollination
syndrome, (6)~the difference in leaf area, and (7) the difference in
leaf length:width ratio. The difference is measured as either 1:0 for
binary data or Euclidean distance for the continuous traits.

Our estimate of phylogenetic distance is derived from a rate-smoothed
version of the phylogenetic tree presented in
\citet{Valente-etal-2010}. Specifically, using the topology presented
in \citet{Valente-etal-2010}, we estimated branch lengths under a
maximum-likelihood model in PAUP$^*$ using the data used to generate
the tree: DNA sequences from four chloroplast markers (\textit{trn}L
intron, \textit{trn}L-\textit{trn}F spacer, \textit{rps}16, \textit
{atp}B-\textit{rbc}L spacer), two nuclear regions (ITS and \textit
{ncp}GS), and 138
AFLP loci. We smoothed the branch lengths using NPRS in r8s
[\citet{Sanderson-2003}] and calculated pairwise phylogenetic distances
using cophenetic.phylo from APE [\citet{Paradis-etal-2004}].

\section{The symmetric power link family}\label{sec3}
Let us first specify the notation used throughout this paper.
Suppose $y_i \sim\operatorname{Binomial}(p_i, N_i)$, where $p_i$ is the
probability of success for the $i$th observation. Let the design
matrix be $\boldX$ with $x_i=(1, x_{i1},x_{i2},\ldots,x_{ik})'$ the
$i$th row of $\boldX$ and $\boldbeta=(\beta_0, \beta_1,\beta
_2,\ldots,\beta_k)'$
the corresponding regression coefficients. We associate $p_i$
and $x_i$ through a c.d.f. $F$
as follows:
%
%
\begin{equation}
\label{eq1} p_i = F\bigl(x_i'\boldbeta
\bigr),
\end{equation}
where we call $F^{-1}$ the corresponding link function. The
logit, probit, Student $t$ link as well as the cloglog link functions
are common links adopted for the binomial regression models.

Here we propose a general class of flexible link functions based on a
symmetric baseline link function and its mirror reflection in the
following manner. If $F_0^{-1}$ is a baseline link function
with corresponding c.d.f. $F_0$ for which the p.d.f. is symmetric about
zero, we propose the symmetric power link family based on $F$ as
%
%
\begin{equation}
\label{eq2} F(x,r)=F_0^r \biggl(\frac{x}{r}
\biggr)\mathbf{I}_{
(0, 1 ]} (r ) + \bigl[1 - F_0^{{1/r}}
(-rx ) \bigr]\mathbf{I}_{ (1,
+\infty)}(r).
\end{equation}
The intuition for the development of (\ref{eq2}) is to utilize the
fact that $F_0^r(x)$ is a valid c.d.f. and it achieves flexible
left skewness when $r < 1$, while the same property holds for its
mirror reflection $1-F_0^{{1}/{r}}(-x)$ with skewness being
in the opposite direction. By combining the two in one single family
of link functions, we could achieve flexibility in positive as well
as negative skewness symmetrically with respect to the baseline
link. Also, we are scaling $x$ by the same parameter $r$ in the
formulation to prevent the mode of the p.d.f. to be too far away from
zero. Clearly, by introducing an additional parameter $r$ in
(\ref{eq2}), the skewness of the symmetric power link family can be
adjusted from its baseline to achieve more flexibility in modeling
asymmetric data.

One immediate observation in (\ref{eq2}) is that $F(x, 1) =
F_0(x)$, so the proposed family includes the baseline c.d.f.
$F_0$ as a special case. Also, considering the fact that
$F_0$ is symmetric, the proposed symmetric power link family is
continuous at the break point $r=1$, since
%
%
\begin{equation}
\label{eq19} \lim_{r\to1_{+}}F(x, r)=1-F_0(-x)=F_0(x)=F(x,1).
\end{equation}
As we will be dealing with introduction of flexible skewness into the
link function, we specify our measurement of skewness here. We adopt
Arnold and Groeneveld's (\citeyear{arnold1995measuring}) skewness
measure with respect to the mode here. Under certain conditions, the
skewness of a random variable $X$ is defined as $\gamma_M=1-2F(M_x)$,
where $F(\cdot)$ is the c.d.f. of $X$ with corresponding mode $M_x$. By
definition, the skewness is between $-1$ and $1$, with $0$ indicating
symmetry. In (\ref{eq2}), it follows directly that $F(x, r) = 1 - F(-x,
\frac{1}{r})$. In other words, the p.d.f. of the symmetric power family
with power parameter $r$ is the mirror image of the p.d.f. with power
parameter $\frac{1}{r}$. This implies that if the skewness of $F(x,r)$
is $\xi$, then the skewness of $F(x,\frac{1}{r})$ will be $-\xi$. Here,
by combining the power of a standard symmetric link distribution
function and its reflection in one single\vadjust{\goodbreak} link, we can accommodate
flexible skewness in both directions simultaneously, while retaining
the desirable property of having the standard baseline link function as
a special case. We propose three symmetric power link function families
based on different baseline link functions as follows.

\subsection{The symmetric power logit (splogit) link family} 
\begin{figure}

\includegraphics{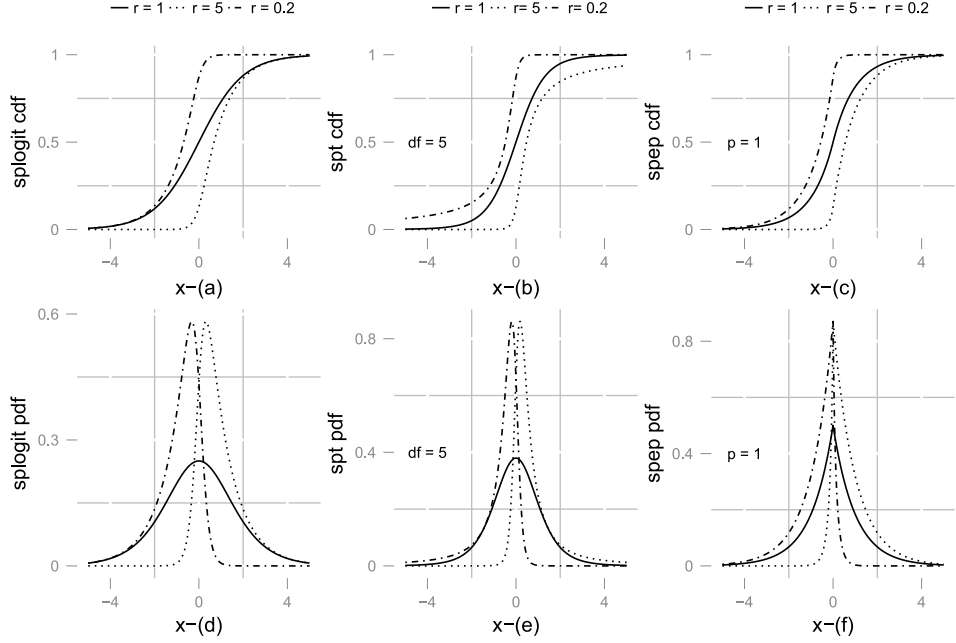}

\caption{Symmetric power link c.d.f. and p.d.f. for different value of r
under logit, Student $t$ and exponential power baseline c.d.f.
functions.} \label{fig1}
\end{figure}

\begin{figure}

\includegraphics{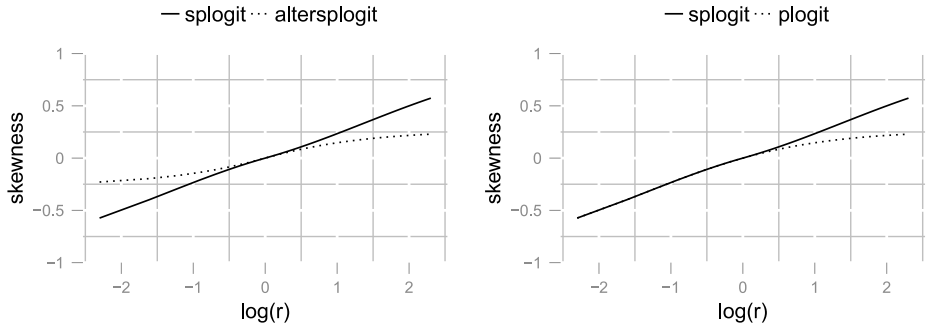}

\caption{Skewness range of splogit against plogit and altersplogit as
$\operatorname{log}(r)$ varies. The
possible skewness ranges from $-1$ to 1 under the definition of
Arnold and Groeneveld (\citeyear{arnold1995measuring}).} \label{fig2}
\end{figure}

If we choose
$F_0$ to follow the logistic distribution with location 0 and
scale 1, then we call $F(x,r)$ defined in (\ref{eq2}) the
symmetric power logit (splogit) family, and we call the corresponding
link function the splogit link. The skewness of the splogit
distribution can be found
analytically as $\gamma_M=1-2 (\frac{r}{r+1} )^r$ for $0 < r
< 1$, and $\gamma_M=2 (\frac{1}{r+1} )^{1/r} - 1$ for $r >
1$. As a result, it is negatively skewed when $0 < r < 1$, positively
skewed when $r > 1$, and reduces to the symmetric logit link when
$r = 1$. Figure~\ref{fig1}(a) and (d) shows the p.d.f. and c.d.f. corresponding
%
%
to the splogit link with $r=0.2, 1, 5$, respectively. It is clear
that as the power parameter $r$ varies, so does the approaching rate
to 0 and 1 for the splogit link. The range of skewness provided by the
splogit family is unlimited, reaching $-1$ and 1, respectively, as $r$
tends to 0 and $+\infty$ [see Figure~\ref{fig2}(a)].

\subsection{The symmetric power $t$ (spt) link family}
Many authors have suggested using a Student $t$ link (degrees of
freedom denoted as $\nu$) as an alternative to the logit and probit
%
%
links. \citet{MudhGeorrema1978} show that the Student $t$ link\vadjust{\goodbreak}
with 9
degrees of freedom has the same kurtosis as the logistic distribution.
\citet{AlbeChibbaye1993} suggest using the $t$ distribution with 8
degrees of freedom and provide a full implementation in a Bayesian
framework. \citet{liu2004robit} proposes the robit model which
uses the
Student $t$ distribution with known or unknown degrees of freedom as
the link function and shows that it is a robust alternative to the
logit and probit model. It is widely known that the Student $t$ link
with large degrees of freedom approximates the probit link. Now, by
bringing in the power parameter $r$ in the sense of (\ref{eq2}), we
can add more flexible skewness in the class of Student $t$ link
functions. We call this new class of link functions the symmetric power
$t$ (spt) family. Similarly, in Figure~\ref{fig1}(b) and (e) we see
the p.d.f. and c.d.f. for the spt link with 5 degrees of freedom for
different values of $r$. Clearly, the spt link family allows us to
adjust both the skewness of the distribution and the heaviness of the
tails by varying $r$ and $\nu$, therefore accounting for an extremely
rich class of link functions. In Arnold and Groeneveld's (\citeyear{arnold1995measuring}) sense,
the closed form expression of the spt link skewness is not available,
but numerically we can show the skewness is quite flexible with small
to medium $\nu$ but becomes more restricted as $\nu$ increases. Notice
that when $r$ is fixed to be positive, the symmetric power probit
distribution is very similar to the power normal distribution proposed
by \citet{gupta2008analyzing}.

\subsection{The symmetric power exponential power (spep) link family}
The exponential power (ep) distribution was first introduced by
\citet{subbotin1923law}. The ep distribution is symmetric with
density
%
%
\begin{equation}
\label{eq3} f(x;\mu,\sigma,p)=c^{-1}\operatorname{exp} \biggl(-
\frac{|z|^{p}}{p} \biggr),
\end{equation}
where $-\infty<x<+\infty, \sigma>0, p\geq1,
z= (x-\mu)/\sigma$, and $c=2\sigma
p^{1/p-1}\Gamma(1/p )$. Clearly, normal distribution is a
special case with $p=2$, and heavier tail distribution can be
obtained as we set $p$ to be less than 2. Also, the Laplace distribution
is another special case when $p=1$. If we set $\mu=0$ and
$\sigma=1$, the ep distribution becomes symmetric about zero and
with flexible tail properties as $p$ varies. If we set the ep as our
baseline link function $F_0$, we end up with the symmetric power
exponential power (spep) link family [see Figure~\ref{fig1}(c) and
(f) for the corresponding p.d.f. and c.d.f. for $p=1$ at different values
of $r$]. Here we restrict $p$ to be within the range of $[1,2]$ for
our proposed spep link family since the skewness of the p.d.f. becomes
restricted for $p>2$, that is, with a thinner tail than normal
distributions. However, even with this restriction, the spep link
family still provides extremely flexible range of skewness and
adjustment of tail behavior in one single family of link functions.

\subsection{Comparison with other power link}
As discussed in Section~\ref{sec1}, many authors have proposed to
bring in a power parameter to allow more flexible skewness in the
link function. Again, let $F_0^{-1}$ be the baseline link; the traditional
power link family is defined by
%
%
\begin{equation}
\label{eq4} H(x,r)=F_0^r(x).
\end{equation}
Also, by choosing a different tail in (\ref{eq2}), there is an
alternative way of constructing the symmetric power link given by
choosing the other side of the tail as follows:
%
%
\begin{equation}
\label{altersp} F^*(x,r)= \bigl[1 - F_0^{{1}/{r}} (-rx ) \bigr]
\mathbf{I}_{ (0, 1 ]}(r) + F_0^r \biggl(
\frac{x}{r} \biggr)\mathbf{I}_{ (1,
+\infty)}(r).
\end{equation}
Here, adopting the logit baseline for all three, we compare model
(\ref{eq2}) (splogit) with model (\ref{eq4}) (plogit) and (\ref
{altersp}) (altersplogit) to
illustrate the advantage of our proposed symmetric power link in terms
of skewness range. Adopting the other baseline link discussed above will
lead to similar results.

The advantage of skewness range of splogit is
illustrated in Figure~\ref{fig2}. Comparing with plogit, when
$\operatorname{log}(r)$ is negative, the skewness of splogit and plogit is
exactly the same, which is due to the fact that the formulation of
splogit follows closely as plogit when $0<r\leq1$. However, on the
other side,
the skewness of splogit reaches~1 as $\operatorname{log}(r)$ goes to infinity, which has a clear advantage
over plogit with
a skewness limit of 0.264. Similar comparison reveals that by choosing
the appropriate side, splogit has skewness
advantage on both tails over the altersplogit link. A more flexible
skewness means that the probability of success under the splogit link
can approach 0 (or 1)
in a rate that can never be achieved under the plogit or altersplogit
link, which makes it more flexible in dealing with
skewed data, as we will show later in the simulation study.

\section{The prior and posterior proprieties}\label{sec4}
We adopt the following class of prior distributions on our proposed
symmetric power link family and investigate its posterior
proprieties. For regression coefficient $\bolds{\beta}$,\vadjust{\goodbreak} we adopt
the usual uniform prior, that is, $\pi(\boldbeta)\propto1$. For the
power parameter $r$, we adopt a proper gamma prior $\pi(r)$ with mean
one and reasonably large variance. If we denote the likelihood of the
model to be $\mathbf{L}(\boldbeta, r|\boldy)$, consequently, the joint
posterior density of our regression model becomes
%
%
\begin{eqnarray}
\label{eq5} \pi(\boldbeta,r|\boldy)&\propto&\mathbf{L}(\boldbeta
,r|\boldy)\pi(
\boldbeta)\pi(r)
\nonumber\\[-8pt]\\[-8pt]
&\propto&\prod_{i=1}^n\bigl[F
\bigl(x_i'\boldbeta,r\bigr)\bigr]^{y_i}
\bigl[1-F\bigl(x_i'\boldbeta,r\bigr)
\bigr]^{1-y_i}\pi(r).
\nonumber
\end{eqnarray}
Clearly, the posterior distribution is proper if and only if
%
%
\begin{equation}
\label{eq6} \int_{\Re^+}\int_{\Re^k}\mathbf{L}(
\boldbeta,r|\boldy)\pi(r)\,d\boldbeta\,dr < \infty.
\end{equation}
Notice that when $r$ is a point mass at 1, $F(x, r)$ becomes the
baseline link function $F_0$. Our goal is to investigate whether
the introduction of a power parameter $r$ would change the posterior
propriety compared to merely adopting the symmetric baseline function
$F_0$. Here we let $\mathbf{L}_0(\boldbeta|y)$ denote the
likelihood under the corresponding baseline link $F_0$.

%
\begin{theorem}\label{theo1}
If under the baseline link $F_0$ we have
%
%
\begin{equation}
\label{eq7} \int_{\Re^k}\mathbf{L}_0(\bolds{\beta}|
\mathbf{y})\,d\bolds{\beta} < \infty,
\end{equation}
then the posterior under the corresponding power link
$\pi(\boldbeta,r|\boldy)$ is also proper, that is,
%
%
\begin{equation}
\label{eq8} \int_{\Re^+}\int_{\Re^k}\mathbf{L}(
\bolds{\beta},r|\mathbf{y})\pi(r)\,d\bolds{\beta}\,dr < \infty.
\end{equation}
\end{theorem}

Theorem \ref{theo1} states that by introducing an additional power
parameter $r$ in the sense of (\ref{eq2}), the posterior propriety
under the uniform $\boldbeta$ prior is unchanged with a proper prior for
$r$. \citet{chen2000propriety} studied the conditions for the
propriety of the posterior distribution under general link
functions. The following theorem resolves the posterior proprieties
of the three proposed symmetric power link families under uniform
$\boldbeta$ priors. For the spt and spep link families, throughout this
paper we adopt proper priors on $\nu$ and $p$.

%
\begin{theorem}\label{theo2}
Let $\tau_i=1$ if
$y_i=0$ and $\tau_i=-1$ if $y_i=1$, and define $\boldX^*$ to be the
matrix with the $i$th row $\tau_ix_i'$. Suppose the following
conditions hold:

\begin{longlist}
\item
The design matrix $\boldX$ is of full column rank.

\item There is a positive vector $\mathbf{a}=(a_1, a_2,\ldots,
a_n)'\in\Re^n$ such that \mbox{$\mathbf{a}'\boldX^*=0$}.
\end{longlist}
\end{theorem}

Then the proposed splogit and spep links lead to proper posteriors
under the above prior setup, while the same result also holds for
the spt family with degrees of freedom $\nu>k$, where $k$ is the number
of columns of $\boldX$.

\section{Spatial random effects}\label{sec10}
Geological and climatic features vary greatly across the CFR. For
example, the climate in the western part of CFR is Mediterranean with
rainfall concentrated in the winter months, while the climate in the
eastern CFR is more aseasonal. While such climate features could
affect the pattern of \textit{Protea} co-occurrence, we are primarily
interested in identifying biological features that influence
co-occurrence. Thus, we add spatial random effects in the model to
account for all latent, unmeasured environmental effects that are
spatially structured. Here we use the intrinsically conditionally
autoregressive (ICAR) model [\citet{besag1993spatial}] to capture these
spatial effects. The ICAR model has gained increasing usage in the
past two decades due to its convenient implementation in the context
of Gibbs sampling for fitting hierarchical spatial models
[\citet{banerjee2004hierarchical}].

In the \textit{Protea} co-occurrence context we suppose that there are
$K$ parent watersheds and that the proximity matrix $A$ is defined
by $A_{ij}=1$ if watersheds $i$ and $j$ are adjacent and $A_{ij}=0$
otherwise. Following notation in Section~\ref{sec3}, the number of
co-occurrence between species $l$ and $m$ within watershed $k$ is
modeled as
%
%
\begin{eqnarray}
\label{eq14} y_{lmk} &\sim& \operatorname{Binomial}(p_{lmk},
N_k),
\\
p_{lm} &=& F\bigl(x_{lm}'\boldbeta+
w_k\bigr),
\end{eqnarray}
where $N_k$ is the number of smaller watersheds within parent
watershed $k$, and $w_k$ is the spatial random effect associated with
watershed $k$, where $k=1,2,\ldots,K$. At the next stage, the spatial
random effects $(w_1,w_2,\ldots,w_K)$ follow the ICAR model, that is,
%
%
\begin{equation}
\label{eq15} p(w_1,w_2,\ldots,w_K)\propto
\operatorname{exp} \biggl[-\frac{1}{2\tau^2}\sum_{i\neq
j}A_{ij}
(w_i-w_j )^2 \biggr].
\end{equation}
Clearly, (\ref{eq15}) is not a proper distribution, so it cannot be
used to model data directly. However, here we use it as a prior
distribution on the second stage of the hierarchical model which
avoids this problem. In addition, to make $\mathbf{w}$ fully
identifiable, we impose the constraint $\sum_{k}w_k=0$.

\section{Computational issues}\label{sec5}
\subsection{Markov chain Monte Carlo (MCMC) sampling}

The posterior distribution given in (\ref{eq5}) is relatively easy
to sample given the standard baseline link c.d.f.~$F_0$. To run the
Gibbs sampler, we subsequently sample from the complete conditional
distributions $[\boldbeta|r,\boldy]$, $[r|\boldbeta,\boldy]$ (also
$[\nu|\boldbeta,r,\boldy]$ if under spt and $[p|\boldbeta,r,\boldy]$
if under spep). Each draw can be done using the Adaptive Rejection
Metropolis algorithm [\citet{GilkBestTanadap1995}] which is
implemented in JAGS [\citet{plummer2003jags}]. Due to the conditional
nature of the ICAR distribution, the Gibbs sampler of spatial random
effects is conveniently constructed and the details are discussed in
\citet{banerjee2004hierarchical}. All the computations in this paper
are done in JAGS or geoBUGS.

\subsection{Covariate effects}
Czado and Santner (\citeyear{CzadSanteffe1992}) pointed out that it is more appropriate to
compare the covariate effects under different link functions with the
estimated probabilities since the estimates of $\boldbeta$ will depend
on the choice of link functions. In view of this, we use the method
suggested by
\citet{ChibJeliinfe2006} to calculate the average effect of
the covariates estimated probabilities. Here we denote the set of all
parameters in our model to be $\boldtheta$. For example, if we want
to estimate the effect of covariate $x_i$, we integrate out parameters
$\boldtheta$ by its MCMC posterior samples, and marginalize out other
covariates $\mathbf{x}_{-i}$ by their empirical distributions to get an
estimate of the predictive distribution
%
%
\begin{equation}
\label{eq9} [p|x_i,\boldy]=\int[p|x_{i},
\mathbf{x}_{-i},\boldtheta,\boldy]\pi(\boldtheta|\boldy)\pi(
\mathbf{x}_{-i})\,d\boldtheta\,d\mathbf{x}_{-i}.
\end{equation}
Then if we compute this estimated probability under two specific
values of $x_i$, for example, under 0 and 1, the difference in the
computed probabilities gives an estimated effect of covariates $x_i$
as it changes from 0 to 1.

\subsection{Bayesian model comparison}
To compare the performance of models under different link functions,
we calculate two summary measures. The first one is the Deviance
Information Criterion (DIC), which balances the fit of a model to the
data with its complexity. The DIC measure is calculated with the posterior
mean of deviance penalized by the effective number of parameters under
the Bayesian framework [\citet{SpieBestCarlvanbaye2002}]. The
other measure we consider here is the logarithm of the
pseudo-marginal likelihood (LPML), which measures the accuracy of
prediction based on leave-one-out cross-validation ideas. The LPML
measure [\citet{ibrahim2005bayesian}] is a summary statistic of the
conditional predictive
ordinate (CPO) criterion [\citet{GelfDeyChanmode1992}]. The model
with the larger LPML indicates better fit of competing
models.

\section{Simulation study}\label{sec6}
Here we conduct three sets of simulation studies. The first one compares
the performance of our proposed symmetric power link function against
some other standard or flexible link functions. The second one
focuses on the performance of splogit link versus plogit link when the
data is generated by a skewed distribution. The third one investigates
specifically how our proposed model performs against other flexible
link functions on a larger scale simulation. To simplify our
simulation, here we consider Bernoulli responses instead of binomial.
Before we go any further, we introduce two other flexible link
functions for comparison purposes.

\citet{StukStukgene1988} proposed the generalized logistic link
family with parameter $\bolds{\alpha} =(\alpha_1, \alpha_2)'$
as follows:
\[
p_i = G\bigl(h_\alpha\bigl(x_i'
\boldbeta\bigr)\bigr),
\]
where $G$ is the c.d.f. of the logistic distribution. When
$x_i'\boldbeta
\geq0$,
\[
h_\alpha\bigl(x_i'\boldbeta\bigr) = \cases{
\alpha_1^{-1} \bigl(\operatorname{exp} \bigl(\alpha_1x_i'
\boldbeta\bigr)-1 \bigr), &\quad$\alpha_1>0$,
\vspace*{1pt}\cr
x_i'
\boldbeta, &\quad$\alpha_1=0$,
\cr
-\alpha_1^{-1}
\operatorname{log} \bigl(1-\alpha_1x_i'\boldbeta
\bigr), &\quad$\alpha_1<0$,}
\]
and for $x_i'\boldbeta\leq0$,
\[
h_\alpha\bigl(x_i'\boldbeta\bigr) = \cases{ -
\alpha_2^{-1} \bigl(\operatorname{exp} \bigl(\alpha_2\bigl|x_i'
\boldbeta\bigr| \bigr)-1 \bigr), &\quad$\alpha_2>0$,
\vspace*{1pt}\cr
x_i'
\boldbeta, &\quad$\alpha_2=0$,
\cr
\alpha_2^{-1}
\operatorname{log} \bigl(1-\alpha_2\bigl|x_i'\boldbeta\bigr|
\bigr), &\quad$\alpha_2<0$.}
\]
Also, \citet{Czadpara1994} proposed another two parameters family
link functions given by specifying
\[
h_\alpha\bigl(x_i'\boldbeta\bigr)= \cases{
\displaystyle \frac{ (x_i'\boldbeta+1 )^{\alpha_1}-1}{\alpha_1}, &\quad$x_i'\boldbeta
\geq0$,
\vspace*{2pt}\cr
\displaystyle -
\frac{ (-x_i'\boldbeta+1 )^{\alpha_2}-1}{\alpha_2}, &\quad
$x_i'\boldbeta< 0$.}
\]

First we compare splogit link with other link functions with detailed
simulation. We generate 2 covariates for our simulation study. We
independently generate one binary covariate $x_1$ with 0 and 1
randomly chosen, and the second covariate $x_2$ is generated
independently from $N(0,3)$. Our vector of covariates is denoted
as $X = (1,x_1,x_2)'$. The true regression coefficient
$\bolds{\beta}=(\beta_0, \beta_1, \beta_2)'$ is set to be
$(0,1,1)'$ for all simulations. With the same value of $X'\beta$, we
carry out our studies under three scenarios based on three true models
as follows:

%
\begin{scenario}\label{scen1}
The binary data are generated from the
symmetric logistic link model with
$F^{-1} (p_i )=\operatorname{log}(p_i/
(1-p_i ) )$.
\end{scenario}

%
\begin{scenario}\label{scen2}
The binary data are generated from the
complementary loglog (cloglog) link with
$F^{-1} (p_i )=\operatorname{log} (-\operatorname{log}(1-p_i ) )$.
It is easily calculated that the skewness of the corresponding
$F$ is $-0.264$, under the definition of \citet{arnold1995measuring}.
\end{scenario}

%
\begin{scenario}\label{scen3}
The binary data are generated from the loglog
link with
$F^{-1} (p_i )=-\operatorname{log} (-\operatorname{log}
(p_i ) )$.
The corresponding $F$ is the mirror reflection of the c.d.f.
corresponding to the cloglog link, and therefore with skewness
$0.264$.
\end{scenario}

\begin{figure}

\includegraphics{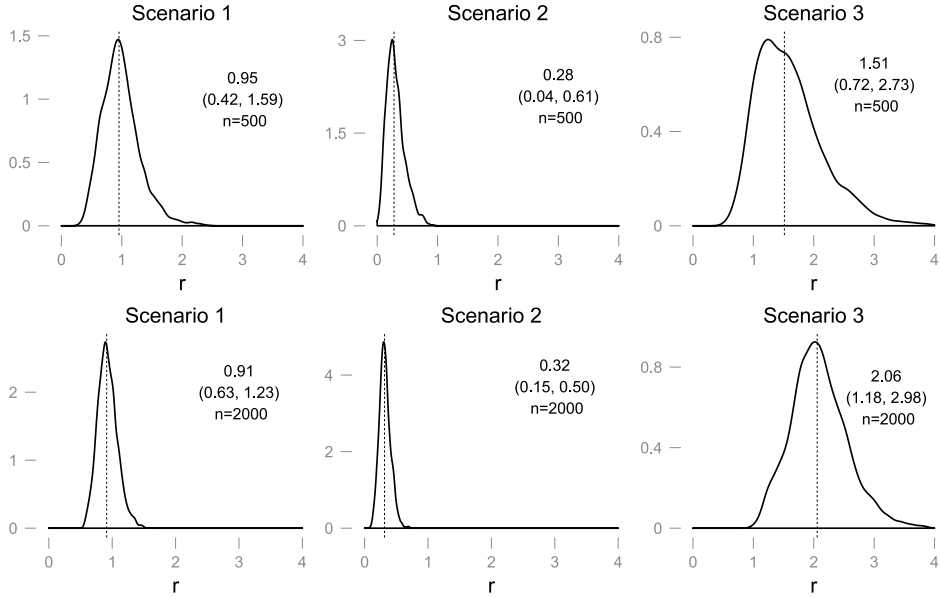}

\caption{Posterior density plot for power parameter $r$ under the
splogit model with sample sizes 500 and 2000. The posterior median and
HPD interval of $r$ are also reported in the plot.} \label{fig4}
\end{figure}

As described in Section~\ref{sec5}, we conduct a fully Bayesian
analysis on the above three simulated data sets. The prior of
$\beta$ is chosen to be $N(0, 10^4)$ and the prior for
$r$ is set to be exponential with parameter 1. In the spt model the
prior for $\nu$ is chosen to be $\operatorname{Gamma}(8,1)$ and for the
spep model
the prior for $p$ is chosen to be $\operatorname{unif}(1,2)$. In the ``Stukel''
and ``Czado'' models, the priors for parameters $a_1$ and $a_2$ are set
to be $N(0, 10^2)$. For each scenario, we repeat the same setting for
two different sample
sizes $N=500$ and $N=2000$ to see how sample size would affect our
inference. After 2000 burn-ins, the models mix pretty well and we
obtain 4000 posterior samples for each parameter. We summarize DIC and
LPML measures in order to make model comparisons. Our simulation
results are summarized in
Table~\ref{tab1} and in Figure~\ref{fig4}.\looseness=1

%
\begin{sidewaystable}
\textwidth=\textheight
\tablewidth=\textwidth
\caption{Posterior covariate effects for $\beta_1$ and $\beta_2$,
posterior median for $\nu$ and $p$ in spt and spep models as well as
model comparisons under sample sizes 500 and 2000. Bold numbers
indicate the corresponding fit is under the true model. Covariate
effects are measured between value 0 and 1. Larger LPML and smaller DIC
indicate a better fit}
\label{tab1}
\begin{tabular*}{\tablewidth}{@{\extracolsep{\fill}}l c c c c d{4.1} c c c c c d{4.1} c c c c c d{4.1}@{}}
\hline
& \multicolumn{17}{c@{}}{\textbf{True model}}\\[-4pt]
& \multicolumn{17}{c@{}}{\hrulefill}\\
\multirow{2}{35pt}[-6.2pt]{\textbf{Fitted model}}
& \multicolumn{5}{c}{\textbf{logit}} && \multicolumn{5}{c}{\textbf{cloglog}}
&& \multicolumn{5}{c@{}}{\textbf{loglog}}\\[-4pt]
& \multicolumn{5}{c}{\hrulefill} && \multicolumn{5}{c}{\hrulefill}
&& \multicolumn{5}{c@{}}{\hrulefill}\\
& $\bolds{\beta_1}$ & $\bolds{\beta_2}$ & $\bolds{\nu/p}$ & \textbf{LPML} & \multicolumn{1}{c}{\textbf{DIC}}
&& $\bolds{\beta_1}$ & $\bolds{\beta_2}$ & $\bolds{\nu/p}$ & \textbf{LPML} & \multicolumn{1}{c}{\textbf{DIC}}
&& $\bolds{\beta_1}$ & $\bolds{\beta_2}$ & $\bolds{\nu/p}$ & \textbf{LPML} & \multicolumn{1}{c@{}}{\textbf{DIC}}\\
\hline\\[-8pt]
& \multicolumn{17}{c@{}}{$\mbox{Sample size} = 500$}\\
[4pt]
logit & \textbf{0.12} & \textbf{0.11} & & \textbf{$\bolds{-}$180.6} & \multicolumn{1}{c}{\hphantom{0}\hspace*{-0.3pt}\textbf
{361.1}} & & 0.15 &
0.06 & & $-$137.6 & 274.7 & & 0.13 & 0.17 & & $-$136.9 & 273.6\\
cloglog &0.10 & 0.15 & & $-$184.3 & 367.4 & & \textbf{0.15} &
\textbf{0.05} & & \textbf{$\bolds{-}$131.1} &
\multicolumn{1}{c}{\hphantom{0}\hspace*{-0.3pt}\textbf{262.1}}& & 0.12 & 0.23 &
& $-$145.1 & 289.5\\
loglog & 0.15 & 0.09 & & $-$189.2& 375.4 & & 0.14 & 0.06 & & $-$161.2 &
317.4 & & \textbf{0.14} & \textbf{0.14} & & \textbf{$\bolds{-}$134.4} &
\multicolumn{1}{c@{}}{\hphantom{0}\hspace*{-0.3pt}\textbf{268.6}}\\
Stukel & 0.11 & 0.09 & & $-$183.4 & 365.8 & & 0.16 & 0.04 & & $-$131.7 &
261.5 & &0.13 & 0.14 & & $-$137.2 & 270.5\\
Czado & 0.11 & 0.10 & & $-$182.4 & 363.9 & & 0.16 & 0.04 & & $-$132.0 &
262.3 & & 0.13 & 0.14 & & $-$136.7 & 269.7 \\
splogit & 0.12 & 0.11 & & $-$181.0 & 359.7 & & 0.15 & 0.05 & & $-$131.5 &
262.1 & & 0.13 & 0.16 & & $-$135.9 & 270.9\\
spt & 0.12 & 0.11 & 7.23 & $-$181.1 & 360.5 & & 0.15 & 0.04 & 6.02 &
$-$131.7 & 261.9 & & 0.13 & 0.16 & 7.43 & $-$136.4 & 271.7\\
spep & 0.11 & 0.11 & 1.41 & $-$181.2 & 360.6 & & 0.15 & 0.05 & 1.23 &
$-$132.6 & 264.0 & & 0.13 & 0.16 & 1.38 & $-$135.9 & 270.9\\
[4pt]
& \multicolumn{17}{c@{}}{$\mbox{Sample size} = 2000$}\\
[4pt]
logit & \textbf{0.13} & \textbf{0.10} & & \textbf{$\bolds{-}$696.6} &
\multicolumn{1}{c}{\hspace*{-0.3pt}\textbf
{1393.1}} & & 0.09 &
0.04 & & $-$491.3 & 982.5 & & 0.11 & 0.16 & & $-$507.0 & 1013.9\\
cloglog &0.13 & 0.14 & & $-$711.6 & 1422.0 & & \textbf{0.09} &
\textbf{0.02} & & \textbf{$\bolds{-}$477.7} &
\multicolumn{1}{c}{\hphantom{0}\hspace*{-0.3pt}\textbf{955.6}} & & 0.11 & 0.22 &
& $-$544.7 & 1088.0\\
loglog & 0.12 & 0.08 & & $-$730.3 & 1457.4 & & 0.09 & 0.05 & & $-$525.8 &
1050.0 & & \textbf{0.11} & \textbf{0.13} & & \textbf{$\bolds{-}$500.6} &
\multicolumn{1}{c@{}}{\hspace*{-0.3pt}\textbf{1000.9}}\\
Stukel & 0.13 & 0.10 & & $-$698.5 & 1396.7 & & 0.10 & 0.02 & & $-$478.9 &
956.6 & & 0.10 & 0.14 & & $-$501.8 & 1003.5\\
Czado & 0.13 & 0.10 & & $-$698.3 & 1396.2 & & 0.10 & 0.02 & & $-$479.0 &
957.1 & & 0.10 & 0.13 & & $-$501.4 & 1002.8 \\
splogit & 0.13 & 0.10 & & $-$696.9 & 1392.9 & & 0.09 & 0.02 & & $-$481.2 &
962.4 & & 0.11 & 0.13 & & $-$501.4 & 1002.2\\
spt & 0.13 & 0.10 & 7.21 & $-$697.5 & 1394.2 & & 0.09 & 0.03 & 9.50 &
$-$483.1 & 966.4 & & 0.11 & 0.13 & 8.54 & $-$502.3 & 1004.1\\
spep & 0.12 & 0.10 & 1.40 & $-$697.5 & 1394.1 & & 0.09 & 0.03 & 1.43 &
$-$478.9 & 956.9 & & 0.10 & 0.15 & 1.34 & $-$501.7 & 1002.4 \\
\hline
\end{tabular*}
\end{sidewaystable}

%

We notice from Figure~\ref{fig4} that under sample sizes of both 500
and 2000, the posterior mean of $r$ in the splogit link is close to 1 when
the true model is logit (symmetric), significantly less than 1 when
the true model is cloglog (left-skewed), and significantly greater
than 1 when the true model is loglog (right skewed). Not surprisingly,
the posterior standard deviation of $r$ becomes considerably smaller
as the sample size increases from 500 to 2000. Also, the performance
of $r$ in the spt and spep links is quite similar to the splogit link shown
in Figure~\ref{fig4}. In conclusion, the power parameter $r$ captures
the skewness of the true model very well.

Table~\ref{tab1} summarizes some other simulation results of the study.
Comparing with standard links (logit, cloglog, loglog), calculated
average covariate effect of $\beta_1$ and $\beta_2$ for the symmetric
power links tends to be much closer to the value under the true model
(bold) than other standard models (logit, cloglog, loglog when it is
not the true model). We also observed that the symmetric power link
provides estimates of covariate effects that are extremely close to
the true model and significantly better than other standard link
functions in terms of LPML and DIC. On the other hand, the symmetric
power link performs extremely close to ``Stukel'' and ``Czado'' models in
terms of both covariate effects and model comparisons. Overall, our
proposed model performs well and proves to be robust enough to handle
various scenarios under simulated data with symmetric, positive and
negative skewness.

We conduct the second simulation study to examine the performance of
the splogit link against the plogit link (\ref{eq4}) and altersplogit
link (\ref{altersp}), when
the data is generated from distributions with various skewness. We simulate
data from the generalized extreme value (gev) distribution with
c.d.f. as follows:
%
%
\begin{equation}
\label{geveq} G(x)=\operatorname{exp} \biggl[- \biggl(1+\xi\frac{x-\mu
}{\sigma
}
\biggr)_{+}^{-{1}/{\xi}} \biggr].
\end{equation}
\citet{WangDeygene2010} propose the gev model as another flexible
link function to model binary response data. The skewness of
(\ref{geveq}) is controlled by the shape parameter $\xi$, which is
calculated as $1-2\operatorname{exp}\{-(1+\xi)\}$ under
Arnold and Groeneveld's (\citeyear{arnold1995measuring}) definition. Notice that $\xi=-0.3$
indicates the skewness is zero. To simulate data from the gev
distribution, we set $F$ to be $G$ with $\mu=0$, $\sigma=1$,
and adopt the same covariates setup as in the first study. We choose
$\xi=-3.3,-0.3,2.7$, respectively, to represent left-skewed, symmetric
and right-skewed data. For each value of $\xi$, we generate 100 data
sets and fit splogit against plogit and altersplogit, respectively, to
compare the average performance of the two. Again, we obtain 4000
posterior samples for each simulation after 2000 burn-in periods.

%
\begin{figure}

\includegraphics{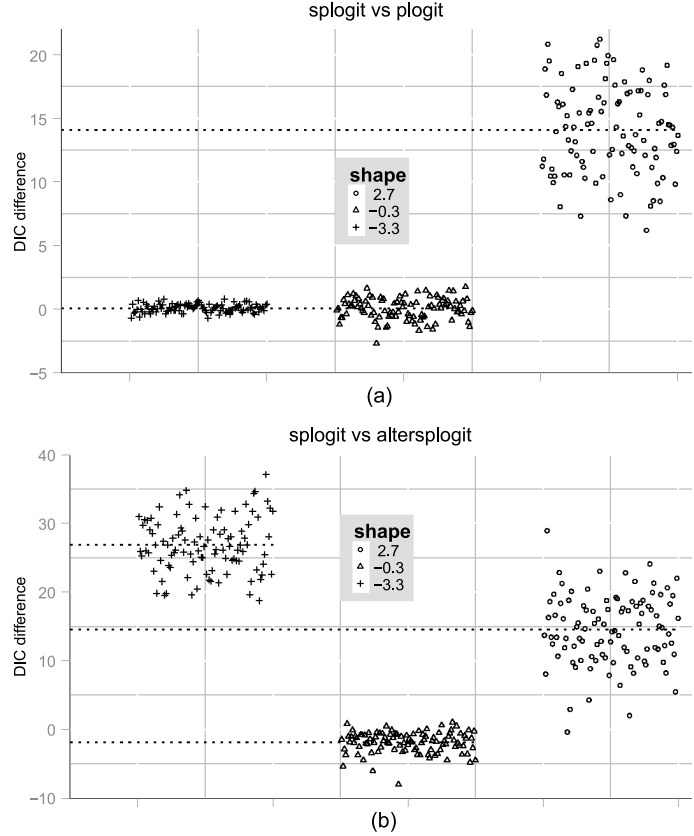}

\caption{Difference of DIC comparing \textup{(a)} splogit and plogit,
\textup{(b)}
splogit and altersplogit when shape
parameter $\xi=-3.3,-0.3,2.7$, respectively. Positive DIC difference indicates
splogit has a better fit. In each case the simulation is repeated
100 times. The dotted line indicates the mean of DIC for the
repetitions.} \label{fig5}
\end{figure}

Figure~\ref{fig5} summarizes the difference of DIC between the
model fits under splogit and plogit (a), splogit and altersplogit (b)
of 100 replicates for each
simulation. For (a), the advantage of using the average DIC of splogit
is not obvious when $\xi= -3.3, -0.3$, but becomes positive at $\xi
= 2.7$. For (b), the average DIC advantage of splogit is positive at
$\xi=-3.3, 2.7$, but ignorable at $\xi=-0.3$.
This is exactly what we expected by looking at Figure~\ref{fig2} in
that the splogit has skewness advantage over plogit when the data is
left skewed, and has skewness advantage over altersplogit in both
skewness directions.

In the third study we compare the performance of the splogit model
against gev, ``Stukel'' and ``Czado'' models on a larger scale, while only
focusing on model comparisons. Here, by larger scale we mean repeating
the fitting of all 4 models 200 times under different true models, and
record the best performance each time according to the DIC measure. To
simplify the comparison, we pick sample size $N=200$ and generate one
continuous covariate from standard normal distribution. For each of the
200 simulations, the true value of $\bolds{\beta}=(\beta_0,
\beta_1)'$ is generated from $N(1, 0.1^2)$. The prior setups of
splogit, ``Stukel'' and ``Czado'' are the same as before, while the prior
of $\xi$ in the gev model is set to be $\operatorname{uniform}(-1, 1)$. The
simulation scheme is also similar to the first study, as we set the
true model to be from logit, cloglog and loglog, then we fit splogit,
gev, ``Stukel'' and ``Czado'' models, respectively, to find the
percentage of
best performance in terms of DIC.

%
\begin{table}
\tablewidth=288pt
\caption{Percentage of best performance among splogit, gev, ``Stukel''
and ``Czado'' out of 200 simulations. The best performance is determined
each time by the lowest~DIC value}
\label{dicperform}
\begin{tabular*}{\tablewidth}{@{\extracolsep{\fill}}l c c c c@{}}
\hline
& \multicolumn{4}{c}{\textbf{\% Lowest DIC}} \\
\hline
True model & \textbf{splogit} & gev & Stukel & Czado \\
logit & \textbf{49.5\%} & 31.0\%\hspace*{1.5pt} & 19.5\% & 0\%\hphantom{0.} \\
cloglog & 26.5\%\hspace*{1pt} & \textbf{62.5\%} & \hphantom{0}8.5\% &2.5\% \\
loglog & \textbf{63.0\%} & \hphantom{0}9.0\%\hspace*{1.5pt} & 24.0\% & 4.0\% \\
\hline
\end{tabular*}
\end{table}

In Table~\ref{dicperform} we clearly see the advantage of the proposed
splogit link over other link functions. The splogit link model performs
the best when the true model is logit and loglog at 49.5\% and 63.0\%,
respectively, where the gev model and ``Stukel'' model come as distant
second places with 31.0\% and 24.0\%. The gev model outperforms splogit
when the true model is cloglog, however, this is expected since cloglog
is a special case of a gev model when $\xi= 0$. Nevertheless, in the
cloglog case, the splogit model performs better than ``Stukel'' and
``Czado'' models at 26.5\%. Overall, we see the robustness of the
proposed splogit link against other flexible link functions.

\section{Data analysis}\label{sec7}
Here we apply the model described in Section~\ref{sec3} on the
\textit{Protea} species co-occurrence data. The data is provided as
supplementary material [\citet{supplement2013}]. As discussed earlier,
we include phylogenetic distance (GD), fire survival strategy (FSS),
plant height, month of maximum flowering (MMF), pollination syndrome,
specific leaf area (SLA) and leaf length width ratio (LWR) as factors
in the model and prevalence probability (with a logit transformation)
as a covariate. Among them, FSS, MMF and pollination are binary and the
rest are continuous. Notice that in order to model the species
co-occurrence, \citet{palmgren1989regression} proposed a method of
running two logistic regressions on two species separately and related
the two with a regression on odds ratio. While it is no problem to
%
%
\begin{sidewaystable}
\tabcolsep=4pt
\textwidth=\textheight
\tablewidth=\textwidth
\caption{Posterior median, HPD interval and DIC measure under splogit,
plogit, gev, logit models of the Protea co-occurrence data. The first
half of the table models $P(y=1)$, while the second half models
$P(y=0)$. Bold numbers indicate significant factors. Smaller DIC
indicates better fit}
\label{tab2}
\begin{tabular*}{\tablewidth}{@{\extracolsep{\fill}}l r c c r c c r c c r c@{}}
\hline
& \multicolumn{2}{c}{\textbf{splogit}}& &\multicolumn{2}{c}{\textbf{plogit}}& &
\multicolumn{2}{c}{\textbf{logit}} & &
\multicolumn{2}{c@{}}{\textbf{gev}}\\[-4pt]
& \multicolumn{2}{c}{\hrulefill}& &\multicolumn{2}{c}{\hrulefill}& &
\multicolumn{2}{c}{\hrulefill} & & \multicolumn{2}{c@{}}{\hrulefill}\\
\textbf{Variables} & \multicolumn{1}{c}{\textbf{Median}\hspace*{-1pt}} & \textbf{HPD interval}
& & \multicolumn{1}{c}{\textbf{Median}\hspace*{-1pt}} & \textbf{HPD interval} & & \multicolumn{1}{c}{\textbf{Median}\hspace*{-1pt}}
& \textbf{HPD interval} & & \multicolumn{1}{c}{\textbf{Median}\hspace*{-1pt}} & \multicolumn{1}{c@{}}{\textbf{HPD interval}}\\
\hline\\[-8pt]
& \multicolumn{11}{c@{}}{{Modeling $P(y=1)$}}\\
[4pt]
Intercept & $-$0.294 &($-$0.354, $-$0.235) & &$-$0.300 &($-$0.349, $-$0.248) &
&0.457 & (0.407, 0.509) & & 0.019 & ($-$0.023, 0.063)\\
GD & \textbf{0.019}& (0.005, 0.033)& &\textbf{0.019} &(0.004, 0.032)
&&\textbf{0.032} & (0.013, 0.052) & & \textbf{0.020} & (0.002,
0.036)\\
FSS &0.025 &($-$0.004, 0.059) && 0.025&($-$0.002, 0.050) && \textbf
{0.053}&(0.010, 0.098) & & 0.027 & ($-$0.007, 0.059)\\
Height & \textbf{$\bolds{-}$0.023}& ($-$0.038, $-$0.009) && \textbf
{$\bolds{-}$0.023}&($-$0.038, $-$0.009) &&\textbf{$\bolds{-}$0.032} &($-$0.052, $-$0.012) & &
\textbf{$\bolds{-}$0.026} & ($-$0.043, $-$0.008)\\
MMF & 0.013&($-$0.026, 0.052) && 0.014& ($-$0.026, 0.050)&& 0.037& ($-$0.014,
0.092) & & 0.014 & ($-$0.031, 0.058)\\
Pollination & $-$0.029 & ($-$0.060, 0.004) && $-$0.029 &($-$0.057, 0.001)
&&$-$0.030 & ($-$0.073, 0.011) && \textbf{$\bolds{-}$0.038} & ($-$0.074, $-$0.003)\\
SLA & $-$0.003 & ($-$0.017, 0.011) &&$-$0.003 &($-$0.018, 0.010) && $-$0.006
&($-$0.024, 0.012) && $-$0.006 & ($-$0.022, 0.010)\\
LWR & \textbf{$\bolds{-}$0.027} & ($-$0.044, $-$0.012) &&\textbf{$\bolds{-}$0.027} &($-$0.043,
$-$0.011) &&\textbf{$\bolds{-}$0.037} &($-$0.059, $-$0.017) && \textbf{$\bolds{-}$0.031} &
($-$0.051, $-$0.013)\\
Prevalence & \textbf{0.844}&(0.815, 0.873) & & \textbf{0.842}&
(0.818, 0.866)& & \textbf{1.146}& (1.125, 1.164) && \textbf{0.991} &
(0.963, 1.016)\\[2pt]
$r/\xi$ & \textbf{0.380}& (0.331, 0.0431)& &\textbf{0.376} &(0.336,
0.423) & & & && \textbf{0.042}&(0.020, 0.062)\\[2pt]
DIC & \multicolumn{2}{c}{23\mbox{,}002.0} & & \multicolumn{2}{c}{23\mbox
{,}001.0} & &
\multicolumn{2}{c}{23\mbox{,}335.2} & & \multicolumn{2}{c}{23\mbox
{,}006.3}\\
[4pt]
& \multicolumn{11}{c@{}}{{Modeling $P(y=0)$}}\\
[4pt]
Intercept & 0.295 & (0.249, 0.333)& &0.313 &(0.293, 0.327) & &$-$0.456
&($-$0.508, $-$0.406) && $-$0.579 & ($-$0.609, $-$0.547)\\
GD & \textbf{$\bolds{-}$0.019}& ($-$0.033, $-$0.003)& &\textbf{$\bolds{-}$0.008} &($-$0.014,
$-$0.003) &&\textbf{$\bolds{-}$0.032} &($-$0.051, $-$0.012) && \textbf{$\bolds{-}$0.015} &
($-$0.026, $-$0.003)\\
FSS &$-$0.024 &($-$0.051, 0.005) && \textbf{$\bolds{-}$0.013}&($-$0.026, $-$0.002) &&
\textbf{$\bolds{-}$0.054}&($-$0.097, $-$0.011) && \textbf{$\bolds{-}$0.026} & ($-$0.047,
$-$0.003)\\
Height & \textbf{0.023}& (0.009, 0.037) && \textbf{0.009}&(0.003,
0.015) &&\textbf{0.031} &(0.010, 0.051) && \textbf{0.017} & (0.005,
0.029)\\
MMF & $-$0.013&($-$0.053, 0.023) && $-$0.008& ($-$0.024, 0.008) && $-$0.037&
($-$0.091, 0.016) && 0.006 & ($-$0.024, 0.038)\\
Pollination & 0.030 & ($-$0.001, 0.059) && 0.009 &($-$0.004, 0.021) &&
0.029&($-$0.014, 0.072) && 0.022& ($-$0.002, 0.045)\\
SLA & 0.003 & ($-$0.010, 0.018) &&0.002 &($-$0.004, 0.008) &&
0.006&($-$0.013, 0.024) && 0.006& ($-$0.005, 0.017)\\
LWR & \textbf{0.028} & (0.012, 0.045) &&\textbf{0.011} &(0.004,
0.017) && \textbf{0.037}&(0.015, 0.058) && \textbf{0.022} &(0.009,
0.035)\\
Prevalence & \textbf{$\bolds{-}$0.845}&($-$0.866, $-$0.818) & & \textbf{$\bolds{-}$0.340}&
($-$0.380, $-$0.294)& & \textbf{$\bolds{-}$1.146}& ($-$1.166, $-$1.127) &&
\textbf{$\bolds{-}$0.677} & ($-$0.686, $-$0.669)\\[2pt]
$r/\xi$ & \textbf{2.624}& (2.344, 2.948)& &\textbf{2.998} &(2.693,
3.441) & & & && \textbf{$\bolds{-}$0.639}&($-$0.661, $-$0.619) \\[2pt]
DIC & \multicolumn{2}{c}{23\mbox{,}001.8} & & \multicolumn{2}{c}{23\mbox
{,}069.1} & &
\multicolumn{2}{c}{23\mbox{,}335.0} && \multicolumn{2}{c}{23\mbox
{,}101.5}\\
\hline
\end{tabular*}\vspace*{-5pt}
\end{sidewaystable}
integrate our sp link in the Palmgren model to replace the logit link,
it is not suitable for our particular data since the co-occurrence data
has the traits difference between two species, but not the traits of
two species separately. For simplicity, we only adopt splogit as a
representative of the symmetric power link family, however, adopting
the spt and spep links would lead to similar results. After 10,000
burn-ins the model parameters and spatial random effects mix pretty
well and then another 10,000 samples have been obtained.
Table~\ref{tab2} summarizes the results under different links with ICAR
prior on the spatial random effects. The priors of the regression
coefficients are set to be normal with mean $0$ and variance $10^4$.
The spatial model is realized utilizing GeoBUGS, an add-on to WinBUGS
[\citet{lunn2000winbugs}].

First, let us look at the first half of Table~\ref{tab2}. When we
model the probability of co-occurrence, that is, $P(y=1)$, we see that
for the splogit and plogit models\vadjust{\goodbreak} the estimate of the power parameter is
around 0.38, corresponding to a left-skewed link. Recall that
splogit is equivalent to plogit when $r \leq1$. As a result,
parameter estimates and model comparison criteria for the splogit
and plogit are roughly the same. The gev model with an estimate of $\xi
= 0.042$ also corresponds to a left-skewed model with a DIC slightly
worse than splogit and plogit. Finally, all three flexible links
perform much
better than logit in terms of DIC.

\begin{figure}

\includegraphics{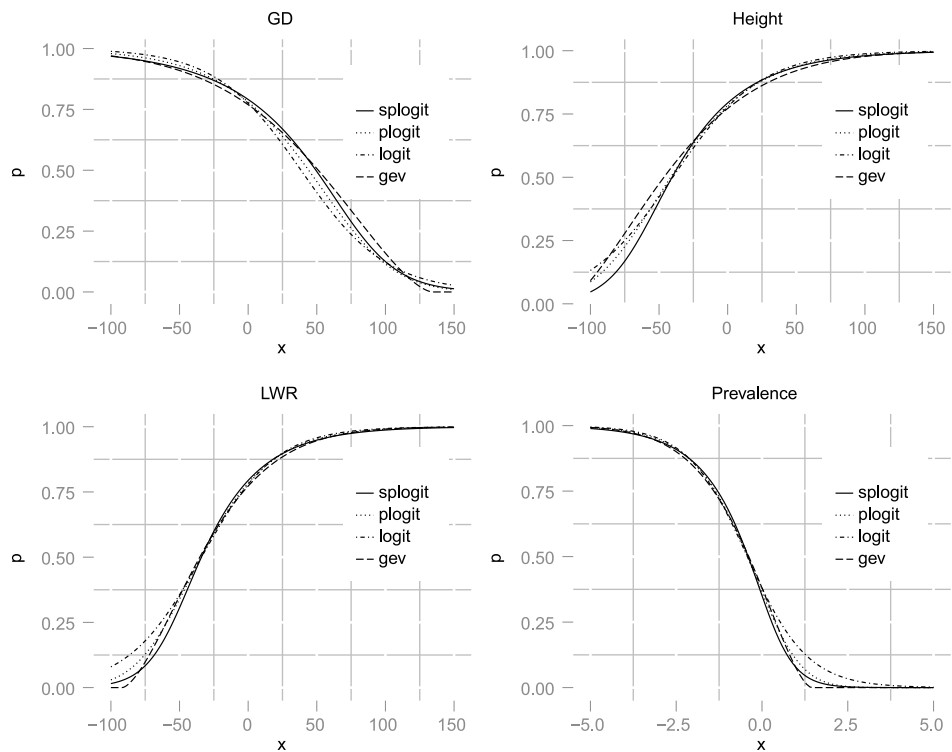}

\caption{Probability curves under splogit, plogit, logit and gev links
as different covariates vary in modeling $P(y=0)$. GD, Height, LWR and
Prevalence are chosen since they are significant in all three models.
Other covariates are fixed to be at the mean and coefficients are fixed
to be at the posterior median.} \label{pcurve}
\end{figure}

In order to show the advantage of the splogit link, the second half of
Table~\ref{tab2} considers the probability of not co-occurring,
$P(y=0)$, instead. We notice that due to the symmetric construction of
splogit, modeling $P(y=0)$ and $P(y=1)$ are essentially the same, the
only change being in the sign of the parameters. Here we
can see that the parameter estimates and model comparison criterion
value of splogit and plogit are different and splogit has a clear edge
in terms of model comparison. These results are consistent with
analytical expectations and our simulations: splogit and plogit are
equivalent when $r\leq1$, but when $r>1$, splogit performs better than
plogit. In the gev model, the estimate of $\xi=-0.639$ now corresponds
to a right-skewed model, and with a DIC of 23,101.5, the gev model fits
worse than splogit and plogit. Again, all three of them fit better than
the standard logit link. Figure~\ref{pcurve} plots
probability curves under splogit, plogit, logit and gev links as
different covariates vary in modeling $P(y=0)$. We see that the curve
under splogit has a more flexible tail behavior that results in a
better fit.

%

In one sense, it is not surprising that ``Prevalence'' is the
predominant influence on the probability of co-occurrence. Let
$n_{ik}$ be the number of small watersheds in which species $i$ is
found in watershed $k$ and $n_k$ be the number of small watersheds in
watershed $k$. Then $p_{ik} = (n_{ik}+1)/(n_k+2)$ is the probability
that species $i$ is found in watershed $k$, and the prevalence of the
species pair $i$ and $j$ in watershed $k$ is $p_{ik}p_{jk}$. In short,
if two species are both common within a watershed, they are likely to
co-occur, and if they are both uncommon, they are unlikely to
co-occur.

In another sense, however, the importance of ``Prevalence'' may be
surprising. It indicates that to a large extent species co-occur or
not as if they were randomly assigned to small watersheds, suggesting
that the biotic factors included in our analysis have relatively
little effect on whether or not they co-occur. Competitive effects, if
they existed, would lead similar species to co-occur less often than
expected [corresponding to negative regression coefficients when
modeling $P(y=1)$]. It may not be surprising that competitive effects
are small in this analysis, since the small watershed scale is much
larger than the scale at which individual plants would compete, but
habitat partitioning among watersheds would lead to the same
pattern. Thus, the small negative coefficients associated with
``Height'' and ``LWR'' (leaf length-width ratio) suggest not only that
competitive effects on the structure of \textit{Protea} communities at
this scale, if any, are small, but also that habitat partitioning has
a similarly small influence on the probability of
co-occurrence.\vadjust{\goodbreak}

We regard ``GD'' (phylogenetic distance) as a proxy for unmeasured
traits that influence co-occurrence, and the positive coefficient on
it may be surprising. Its magnitude is similar to that of the negative
coefficients on ``Height'' and ``LWR,'' and it is small relative to
``Prevalence,'' but the positive sign indicates that closely related
taxa occurring in the same large watershed are more likely to co-occur
within small watersheds than expected by chance. Perhaps this
association reflects some degree of habitat filtering
[\citet{Shipley-etal-2006,weiher1995assembly}] on traits that we
did not
measure in this study.

%
\begin{figure}

\includegraphics{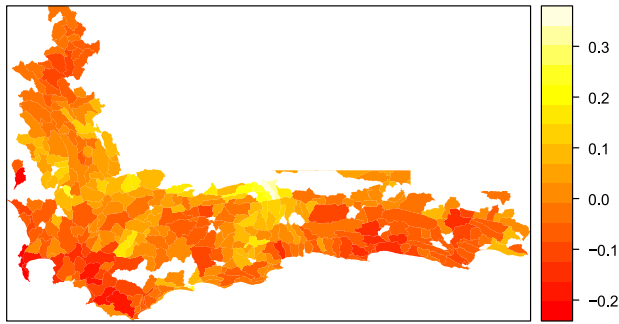}

\caption{Plot of spatial random effects in CFR under splogit fit.}
\label{fig8}
\end{figure}

The spatial clustering effects on co-occurrence
probabilities are also obvious. In Figure~\ref{fig8} we see negative
effects clustered in the east part and southwest corner
of CFR, while positive effects are clustered in the middle part
of the region. This clustering could reflect the interaction of
rainfall (winter rainfall in the west, aseasonal in the east) and
elevation (highest elevations inland, lowest along the coast). In
future studies, we will explore both the patterns of co-occurrence at
different spatial scales and the extent to which climate or other
environmental features are associated with residual spatial variation
in this analysis.

\section{Discussion}\label{sec8}
In this paper we introduced a new family of flexible link functions.
The proposed power link family can accommodate flexible skewness in
both positive as well as negative directions, while retaining the
baseline standard link as a special case. Simulation results show the
proposed link performs well under various skewness scenarios. Also,
the proposed link is computationally straightforward and efficient to
implement. In addition, the power parameter idea illustrated here
may be used to construct new link functions. For example, we could use
an asymmetric link function c.d.f. as our baseline link. Using a power
parameter might make a difference in bringing in desirable flexibility.

One potential problem with the proposed power link is that the power
parameter $r$ influences both the skewness and the mode of the link
function p.d.f. Although this effect has been greatly reduced by
scaling $x$ by $r$ in our model as defined in (\ref{eq2}) and
discussed in Section~\ref{sec3}, the effect still exists with
relatively large $r$ values. One solution might be to adjust the
effect out with calculated mode values, yet it is computationally
expensive especially under pt and pep links when there is no
analytical solution for the mode of the c.d.f. function.

%
\begin{appendix}\label{app}
\section{\texorpdfstring{Proof of Theorem \lowercase{\protect\ref{theo1}}}
{Proof of Theorem 1}} Observing that in our
definition the power link functions $F_r$ naturally split with
respect to $r=1$, we have
\begin{eqnarray*}
\label{eq8} &&\int_{\Re^+}\int_{\Re^k}
\mathbf{L}(\bolds{\beta},r|\mathbf{y})\pi(r)\,d\bolds{\beta}\,dr
\\
&&\qquad=\int_0^1\int_{\Re^k}
\mathbf{L}(\bolds{\beta},r|\mathbf{y})\pi(r)\,d\bolds{\beta}\,dr +\int
_1^\infty\int_{\Re^k}\mathbf{L}(
\bolds{\beta},r|\mathbf{y})\pi(r)\,d\bolds{\beta}\,dr.
\end{eqnarray*}
Clearly, the link function in the latter part is the mirror
reflection of the first part, in other words, $F(x,r) =
1-F(-x,\frac{1}{r})$, therefore, we only need to prove the first
part of the integration is finite. For the baseline link $F_0$
we have
\begin{eqnarray*}
\label{eq9} F_0(x)&=&\int_{\Re}I(u\geq-x)\,d
\bigl(-F _0(-u) \bigr),
\\
1-F_0(x)&=&\int_{\Re}I(u>x)\,dF_0(u).
\end{eqnarray*}
Since $F_0$ is continuous, by Fubini's theorem we have
\begin{eqnarray*}
\label{eqa10} &&\int_{\Re^k}\mathbf{L}_0(\bolds{
\beta}|\mathbf{y})\,d\boldbeta
\\
&&\qquad=\int_{\Re^k}\prod_{i=1}^nF_0
\bigl(x_i'\boldbeta\bigr)^{y_i}
\bigl[1-F_0\bigl(x_i'\boldbeta\bigr)
\bigr]^{1-y_i}\,d\boldbeta
\\
&&\qquad=\int_{\Re^k}\int_{\Re^n}I
\bigl(u_i\geq\tau_ix_i'
\boldbeta, 1\leq i\leq n\bigr)\,d \bigl(\tau_iF_0(
\tau_iu_i) \bigr)\,d\boldbeta
\\
&&\qquad=\int_{\Re^n}\int_{\Re^k}I\bigl(
\boldX^{*\prime}\boldbeta\leq\boldu\bigr)\,d\boldbeta\,d\mathbf{F}_0(
\boldu),
\\
&&\qquad< \infty,
\end{eqnarray*}
where
$\mathbf{F}_0(\boldu)=(\tau_1F_0(\tau_1u_1),\tau_2F
_0(\tau_2u_2),\ldots,\tau_nF_0(\tau_nu_n))$.

Then under the power link, since $|F_0|\leq1$ and $\pi(r)$ is a
proper density, we have
\begin{eqnarray*}
\label{eq11} &&\int_0^1\int
_{\Re^k}\mathbf{L}(\bolds{\beta},r|\mathbf{y})\,d\bolds{\beta}
\pi(r)\,dr
\\
&&\qquad=\int_0^1\int_{\Re^n}\int
_{\Re^k}I\bigl(\boldX^{*\prime}\boldbeta\leq\boldu\bigr)\,d
\boldbeta\,d\mathbf{F}(\boldu,r)\pi(r)\,dr
\\
&&\qquad=\int_0^1\int_{\Re^n}\int
_{\Re^k}I\bigl(\boldX^{*\prime}\boldbeta\leq\boldu\bigr)
\mathbf{F}_0^{r-1} (\boldu)\,d\boldbeta\,d
\mathbf{F}_0(\boldu)\pi(r)\,dr
\\
&&\qquad\leq\int_0^1\int_{\Re^n}
\int_{\Re^k}I\bigl(\boldX^{*\prime}\boldbeta\leq\boldu
\bigr)\,d\mathbf{F}_0(\boldu)\pi(r)\,dr
\\
&&\qquad<\infty.\\[-26pt]
\end{eqnarray*}

\section{\texorpdfstring{Proof of Theorem \lowercase{\protect\ref{theo2}}}
{Proof of Theorem 2}} By Theorem \ref{theo1},
we only
need to prove the theorem for $r=1$. Let $\boldu=(u_1,u_2,\ldots,u_n)'$ be
i.i.d. random variables with distribution function $F_0$.
Now, under the condition listed in Theorem \ref{theo2}, it follows\vadjust{\goodbreak} directly
from Lemma~4.1 of \citet{chen2000propriety} that there exists a
constant K such that
\[
\|\boldbeta\|\leq K\max_{1\leq i\leq n}|u_i|,
\]
whenever
\[
\boldX^{*\prime}\boldbeta\leq\|\boldu\|.
\]
Therefore, following the derivation in Theorem \ref{theo1} under the baseline
link $F_0$, we have
\begin{eqnarray*}
&&\int_{\Re^k}\mathbf{L}_0(\bolds{\beta}|\mathbf
{y})\,d\bolds{\beta}
\\
&&\qquad=\int_{\Re^n}\int_{\Re^k}I\bigl(
\boldX^{*\prime}\boldbeta\leq\boldu\bigr)\,d\boldbeta\,d\mathbf{F}_0(
\boldu)
\\
&&\qquad\leq K\int_{\Re^n}\max_{1\leq i\leq n}|u_i|\,d
\mathbf{F}_0(\boldu)
\\
&&\qquad\leq K\sum_{1\leq i\leq n}E|u_i|^k.
\end{eqnarray*}
Clearly, in the logistic and exponential power cases we have
$E|u_i|^k<\infty$, while in the Student $t$ case the same condition
will hold as long as the degrees of freedom $\nu>k$.
\end{appendix}

\section*{Acknowledgments}

The authors thank Dr. Tony Rebelo and the Protea Atlas project for
providing the \textit{Protea} occurrence data in this paper.

\begin{supplement}
\stitle{\textit{Protea} species co-occurrence data set}
\slink[doi]{10.1214/13-AOAS663SUPP} 
\sdatatype{.zip}
\sfilename{aoas663\_supp.zip}
\sdescription{We provide the \textit{Protea} species co-occurrence data set
used in the data analysis section.}
\end{supplement}

%

\printaddresses

\end{document}